# The nature of 4f electron magnetism in the diluted ferromagnetic Kondo lattice, $CeIr_2B_2$

K. Mukherjee, Kartik K. Iyer, and E. V. Sampathkumaran

*Tata Institute of Fundamental Research, Homi Bhabha Road, Colaba, Mumbai 400005, India*

**Abstract**

We report the physical properties of the series $Ce_{1-x}La_xIr_2B_2$ ($x$ = 0 to 0.9) through magnetization, heat capacity and electrical resistivity measurements as a function of temperature (down to 0.7 K for the latter two measurements). The Curie temperature of $CeIr_2B_2$ (~5 K) is lowered down in temperature due to La substitution, as expected. However, no quantum critical point or 'non Fermi Liquid' behavior was observed even in the dilute limit of $x$=0.9. Interestingly, ferromagnetic ordering persists even for $Ce_{0.1}La_{0.9}Ir_2B_2$, below 0.8 K. Among Ce systems, $CeIr_2B_2$ appears to be one of the compounds in which direct *4f-4f* interaction does not appear to play any role in magnetism, and it is controlled by indirect exchange interaction alone. In this compound, the Kondo effect persists in the ferromagnetic ordered state, as inferred from the entropy data.

# I. Introduction

The Ce-based intermetallics are in the centre-stage of research to probe the competition between the intersite magnetic interaction and on-site Kondo interaction [1, 2]. Several exotic properties have been discovered, attributable to this competition. The strength of hybridization can be tailored by various parameters like composition ($x$), external pressure ($P$) and magnetic field ($H$) [3, 4, 5]. Apart from these factors, it is often recognized that direct $4f$-$4f$ interaction also plays a crucial role in deciding the properties in the case of Ce-systems and it coexists with the indirect Ruderman-Kittel-Kasuya-Yosida (RKKY) exchange interaction [6, 7].

In this article, we focus on the compound $CeIr_2B_2$, which exhibits ferromagnetic (FM) ordering around ($T_C=$) 6 K [8]. Though this behavior was reported more than a decade ago, further extensive investigation is lacking in literature. Hence, it would be interesting to probe how the low temperature properties of Ce evolve with the suppression of $T_C$ due to a change in chemical environment, by the dilution of Ce sub-lattice. For this purpose, we choose to gradually replace Ce with La. It is observed that, while the $T_C$ can be gradually lowered in temperature with the dilution of Ce sub-lattice, there were evidences for the perseverance of magnetic ordering even in $Ce_{0.1}La_{0.9}Ir_2B_2$, implying that indirect RKKY interaction alone, rather than intersite interaction, determines the magnetism of this series.

# II. Experimental details

The compounds, $Ce_{1-x}La_xIr_2B_2$ (x=0.0, 0.3, 0.5, 0.7 and 0.9), were prepared by arc melting stoichiometric amounts of respective high-purity elements in an atmosphere of argon. X-ray diffraction studies established that all the samples are single phase (Fig 1), revealing that all the compounds crystallizes in $CaRh_2B_2$ type orthorhombic structure. A comparison of x-ray diffraction patterns of the parent and La substituted compounds (inset of Fig 1(a)) reveals a gradual shift of diffraction lines with La substitution thereby establishing that La indeed goes to Ce site without precipitating any other phase, within the detection limit (<2%) of this technique.

The *dc* magnetization (*M*) measurements (in the range 1.8-300K) for all specimens were carried out with the help of a commercial superconducting quantum interference device (Quantum Design) magnetometer. We have performed heat capacity (*C*) measurements employing a physical property measurements (PPMS) system (Quantum Design) as a function of temperature (*T*) down to 0.7 K. The electrical resistivity ($\rho$) measurements by standard four probe method, in absence/presence of magnetic fields (upto 100 kOe, *T*=0.7–300K), were performed with the same PPMS. A conducting silver paint was used for making electrical contacts of the leads with the samples.

### III. Results and discussions

#### A. Magnetization

Figure 2(a) shows the temperature dependence of magnetization of $Ce_{1-x}La_xIr_2B_2$ under zero-field-cooled condition, measured in a field of 5 kOe. A sudden change in the curvature of the plot reveals that the $T_C$ gets gradually shifted to a lower temperature with an increase in *x*, falling below 1.8 K for *x*=0.9. This observation is more apparent from *d(M/H)/dT* vs. *T* plots shown in Fig. 2(b). The values of $T_C$ obtained from the temperature at which there is a minimum in this plot for *x*=0.0, 0.3 and 0.5 are ~ 5, 3.7 and 2.5 K respectively. For the extreme compositions it is shifted below 1.8 K. The suppression of $T_C$ in this series as observed experimentally is in accordance with that expected on the basis of indirect exchange interaction. It must be emphasized that if there was a direct 4*f*-4*f* interaction, it would have led to a further suppression of $T_C$. Also, if this compound lies at the right side of the peak in the Doniach's magnetic phase diagram, following increasing unit-cell volume due to La substitution, $T_C$ normalized to Ce concentration should have exhibited an increase. Fig. 2(c) shows the effect of the temperature dependencies of magnetization measured in a field of 5 kOe for various values of external pressure in the vicinity of magnetic transition. The $T_C$ is weakly suppressed under the effect of external pressure. In order to make a comparison with the La substituted compounds, it is desirable to know the bulk modulus of Ce compounds, which is around 600-1000 kbar [9, 10]. From the knowledge of unit-cell volume, it is estimated that the negative pressure exerted by $x$=0.7 compound corresponds to an external pressure of 7 kbar. But it is observed that $T_C$ shifts from ~ 5 to 3.9 K under a pressure of 6 kbar. If the same trend continues by negative chemical pressure, $T_C$ (normalized to Ce concentration) would have gone up by La substitution, in contrast to observations. This implies that the compound lies at the peak of the Doniach's phase diagram [11].

The susceptibility curves of this series in the range 150-300 K were fitted with the modified Curie-Weiss law. The effective moment obtained from the fitting corresponds to that of trivalent Ce for $x$=0.0 and it decreases linearly with increasing $x$. The value of the temperature independent component ($\chi_0$) for all the compositions is ~$10^{-4}$ emu/mol, while the paramagnetic Curie- temperature ($\theta_p$) is around -12 K remaining essentially unchanged across the series, and negative sign of $\theta_p$ for a ferromagnetic Ce alloy imply that the dominant contribution to $\theta_p$ arises from the single-ion Kondo temperature. From figure 2(d), it is inferred that the nature of *M(H)* curve at 1.8 K curves undergoes a change with an increase of dopant concentration. The sharp rise and magnetic hysteresis observed in low fields for $x$=0.0 is gradually suppressed with an increase of $x$ and a smooth curvature of *M(H)* curve is observed for $x$=0.7 for the entire field range. For $x = 0.7$ and 0.9, the ordering temperature lying below 1.8 K is responsible for this observed difference in *M(H)* isotherm. A magnetic moment of 0.85$\mu_B$/formula unit at 50 kOe is observed for $x=0.0$, which linearly decreases with increasing doping (inset of Fig. 2(e)), as expected for a single-ionic nature of the observed magnetic moment. The observed reduced value of the magnetic moment (compared to free $Ce^{3+}$ moment of 2.54 $\mu_B$/formula unit) is attributed to crystal field effect. Also, $T_C$ shifting towards zero does not result in NFL features at low temperatures (e. g. $\chi \sim log$ T) for instance, for the extreme La composition (Fig. 2(f)), as observed often dilute Ce or U alloys in the literature [13].

### B. Heat capacity

The heat capacity behavior of this series as a function of temperature is shown in the figure 3a. For $x=0.0$ an upturn is observed below 6 K in the *C*(T) plot, consistent with magnetic long range ordering around 5 K. This peak shifts to a higher temperature in a field of 50 kOe, implying that the magnetic ordering in this compound is FM (Fig 3(b)). With the increase in doping, the features shift to lower temperatures and move below 1.8 K for $x=0.7$ in zero field. Hence, for this composition, *C* measurement was extended to a further lower temperature using a He-3 insert. A clear peak around 1.4 K is observed in the *C* vs *T* plot of the compound (Fig. 3(c)). For this concentration, a peak is observed in 50 kOe at a higher temperature (4 K), implying that the FM ordering is persisting. For $x=0.9$, a tendency for the formation of a peak is observed below 0.8 K in zero field *C* (Fig. 3(d)) without any peak at higher temperatures. When a field of 50 kOe is applied, *C(T)* develops a peak around 1.6 K, implying that even for this extreme composition of this series, in this study, FM ordering persists. Also, we would like to mention that variation of $C/T$ is not proportional to *log* T, implying an absence of NFL behavior, as inferred from magnetization measurements.

For this series, the magnetic part of the $C$ ($C_{4f}$) was obtained by subtracting the lattice part with $La_2Ir_2B_2$ as the reference for lattice contribution. For this analysis, the difference in Debye temperatures of the series and the reference compound is taken into account using a procedure given is Ref 12. Also the magnetic entropy ($S_{4f}$) was estimated from the 4*f* contribution to $C$. For the parent compound (Fig 3(e)), at $T_C$, $S_{4f}$ = 3.1 J/mol K Ce is ~53% of that expected for the crystal field split of the doublet ground state. This observed lower value of entropy implies a partial screening of Ce magnetic moment by conduction electron spins. Hence this result implies a coexistence of Kondo effect with FM ordering. Similar phenomenon has also been observed for the Ce series, viz: $Ce_yLa_{1-y}Ge_2$ [14]. Under the assumption that the reduction in entropy value is Kondo-derived, an estimate of the single-ion Kondo temperature $T_K$ can be acquired from

$$S_{4f}(T_C) = S_K(T_C/T_K)$$

where, $S_{4f}$ and $S_K$ are the magnetic and Kondo entropy respectively at $T_C$ [15]. According to spin ½ Kondo model, $T_K$ for the parent compound was found to be around 8 K. The linear coefficient of specific heat ($\gamma$) was obtained from $C_{4f}/T$ vs. $T^2$ plot (in the linear $T$ region above $T_C$, which is different for different compositions). The value turns out to be ~ 150 mJ/mol $K^2$ for the parent compound, implying that, $CeIr_2B_2$ is a moderately heavy fermion compound. The value of $\gamma$ decreases and it is ~24 mJ/mol $K^2$ for $x$ = 0.9, as expected for magnetic dilution.

## C. Electrical resistivity

Figure 4 shows $\rho$ plotted as a function to $T$. We have plotted normalized resistivity (at 300K) as the absolute values are not very reliable due to microcracks in the specimens. For the extreme two compositions, $x=0.7$ and $0.9$, $\rho$ measurements were performed up to 0.7 K. For the parent compound, a significant drop of $\rho$ around 5 K is observed and this temperature coincides with the minimum in $d\rho/dT$. This temperature is clearly the magnetic ordering temperature. For all the compositions, the temperature of the drop nearly coincides with the ordering temperature of the respective members. Also, no non-Fermi liquid signature is observed even in the dilute limit, as such a behavior can be expected for an alloy at the quantum critical point, and not for the one near the peak in the Doniach's magnetic phase diagram [11].

## IV. Conclusion

The influence of dilution of Ce sub-lattice on the magnetic behavior of $CeIr_2B_2$ has been investigated. The $T_C$ around 5 K observed for $CeIr_2B_2$ is reduced gradually, with a prevalence finite $T_C$ (below 0.8 K) even for $x=0.9$. If direct $f$-$f$ interaction plays a role, $T_C$ would have decreased more sharply with $x$. Thus this is an interesting Ce system in which magnetization is controlled by indirect exchange interaction alone. And finally, the results establish that $CeIr_2B_2$ is a ferromagnetic Kondo lattice.

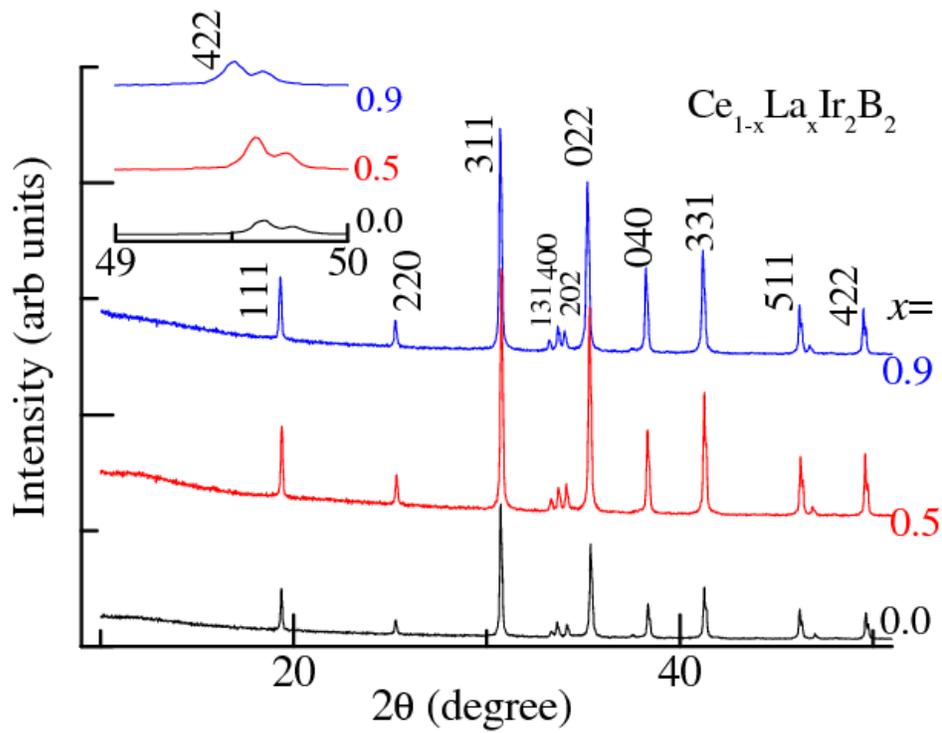

Figure 1:

(color online) (a) X-ray diffraction patterns (Cu $K_\alpha$) below $2\theta=60^\circ$ for the sample series, $Ce_{1-x}La_xIr_2B_2$. The curves are shifted along y axis for clarity. Inset: Same figure showing a gradual shift in diffraction line with increasing concentration at higher angle side.

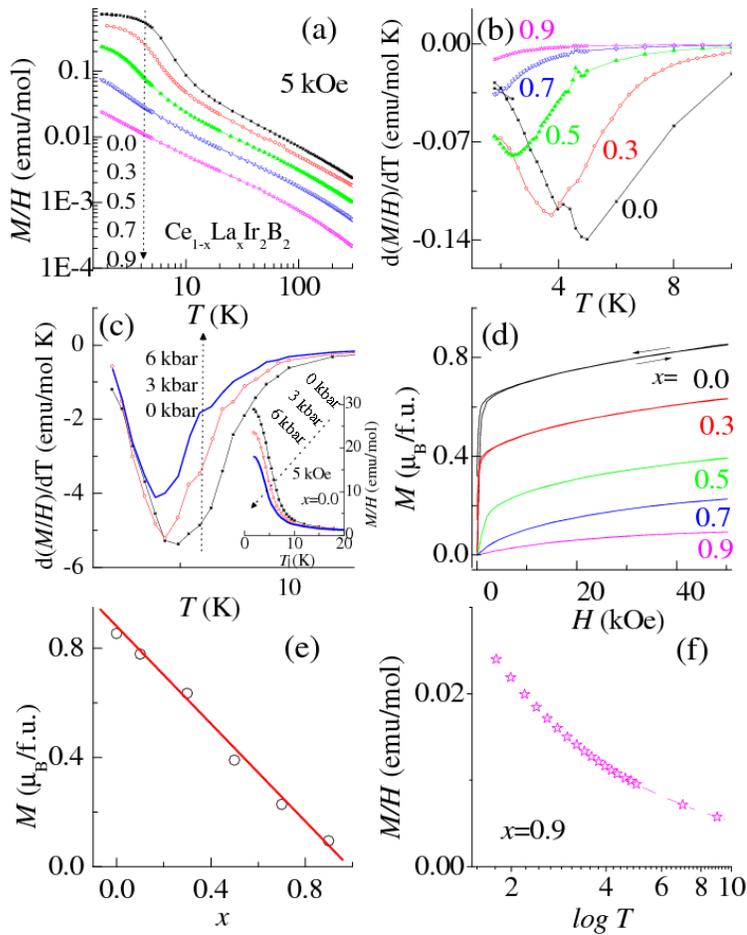

Figure 2:

(color online) (a) Temperature ($T$) dependence (1.8-300 K) of magnetization ($M$) divided by magnetic field ($H$) for $Ce_{1-x}La_xIr_2B_2$ for zero-field-cooled condition. (b) Temperature response of d(M/H)/dT for all the compounds. (c) Temperature response of d($M/H$)/dT for all external pressures. Inset: *M/H verses T* plot under the influence of external pressure. (d) Isothermal magnetization behavior of $Ce_{1-x}La_xIr_2B_2$ at 1.8 K. (e) Magnetization value at 1.8 K at 50 kOe field, plotted as a function of La concentration. A straight line is drawn through data points. (f) *M/H* curve plotted as a function of *log* T for *x*=0.9.

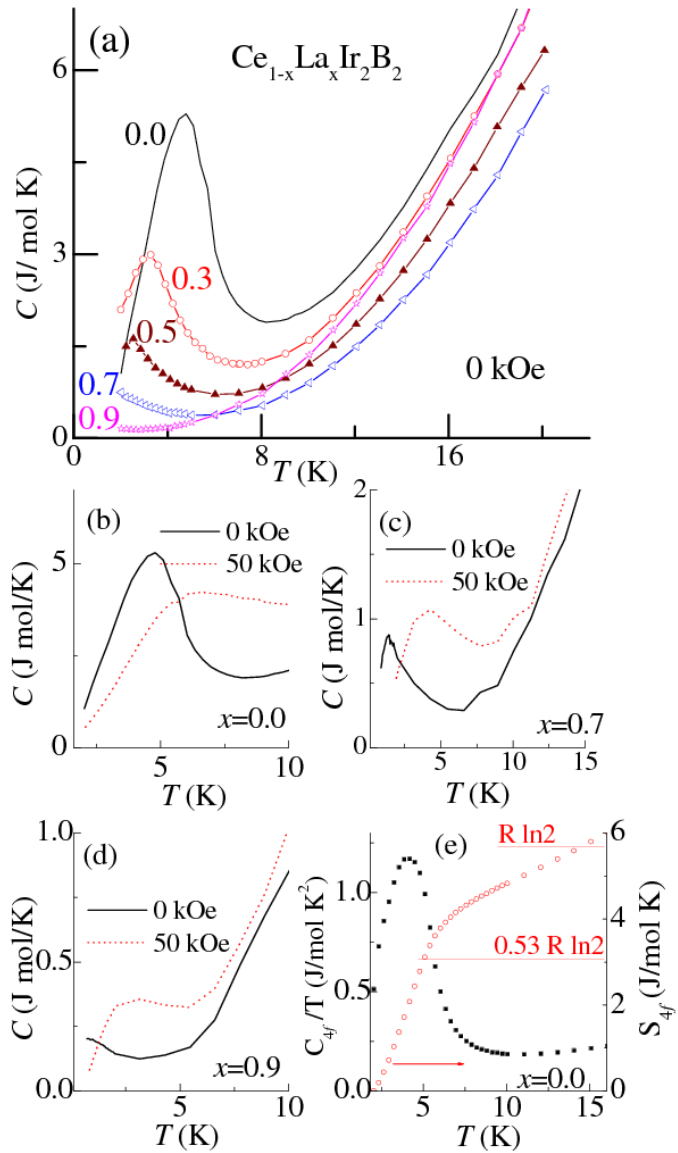

Figure 3:

(color online) (a) Heat capacity of $Ce_{1-x}La_xIr_2B_2$ as a function of temperature in zero field. (b) Heat capacity for $x=0.0$ compound as a function of temperature in zero and 50 kOe field. (c) and (d) Heat capacity for $x=0.7$ and 0.9 compounds as a function of temperature in zero and 50 kOe field down to 0.7 K. (e) $C_{4f}/T$ vs $T$ plot for x=0.0. The magnetic (4$f$ contribution) entropy is presented in the right axis.

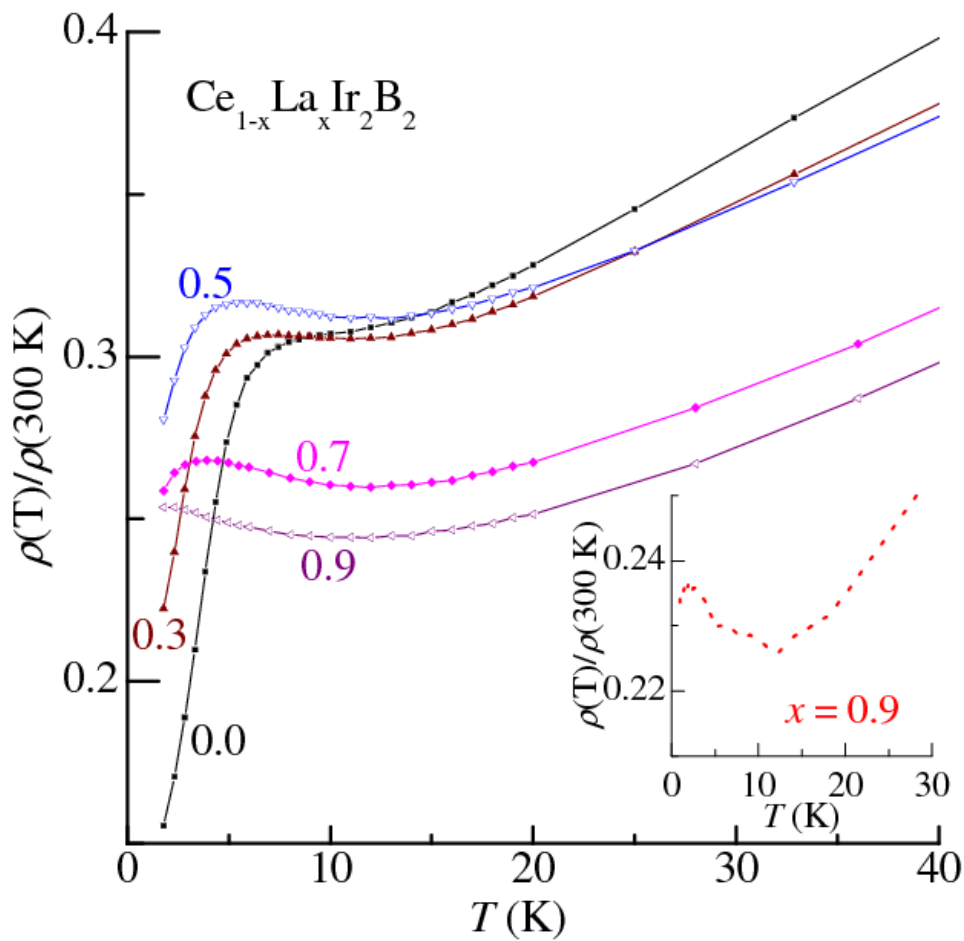

Figure 4:

(color online) Normalized electrical resistivity as a function of temperature in zero field for the compound series $Ce_{1-x}La_xIr_2B_2$. Inset: Normalized resistivity for $x=0.9$ compound measured down 0.7 K.